# Kerberos Authentication in Wireless Sensor Networks

**Qasim Siddique**
**Foundation University, Islamabad, Pakistan**
qasim_1987@hotmail.com

**ABSTRACT** We proposed an authentication mechanism in the wireless sensor network. Sensor network uses the Kerberos authentication scheme for the authentication of bases station in the network. Kerberos provides a centralized authentication server whose function is to authenticate user by providing him the ticket to grant request to the base station. In this paper we have provided architecture for the authentication of base station in the wireless sensor network based on the Kerberos server authentication scheme.

**Introduction**

Wireless sensor network have been used in various application for the monitoring and collection of environmental data. Wireless sensor network are inexpensive consists of large number of sensor nodes. Access to these sensor nodes is organized via a special gateway called base station. This sends queries in the wireless sensor network and retrieves the required data.

Sensor networks provide a powerful solution to many monitoring problems. Nodes in the network may cooperatively monitor physical or environmental conditions, such as temperature, sound, vibration, pressure, motion, and pollution [RSZ03]. However, due to limited resources and short battery life, operations on sensor nodes must be very carefully designed. Moreover, sensors' deployment is not always under direct control and sensors are often under the risk of physical attacks that can affect their security.

In wireless sensor network a base station can't be trusted to identify its users correctly to network services. In particular the following three threads exist.

- A user may gain access to a particular base station and pretend to be another user operating from the base station.
- A may eavesdrop an exchange and use a reply attack to gain entrance to a base station or to interrupt operation.
- A user may alter the network address of a base station so that the request sent from the altered base station to come from the impersonated workstation.

Unauthorized user may gain access to the base station and collect the data that he or she is not authorized to access. Rather that building in elaborates authentication protocol at each sensor node. Kerberos[1] provide a centralized authentication server whose function is to authenticate users to servers and servers to users.

Kerberos relies exclusively on symmetric encryption, making no use of public key encryption. Two versions of the Kerberos[2] are in common use.

- Kerberos 4 [M+88][SNS88]
- Kerberos 5 [SP05]

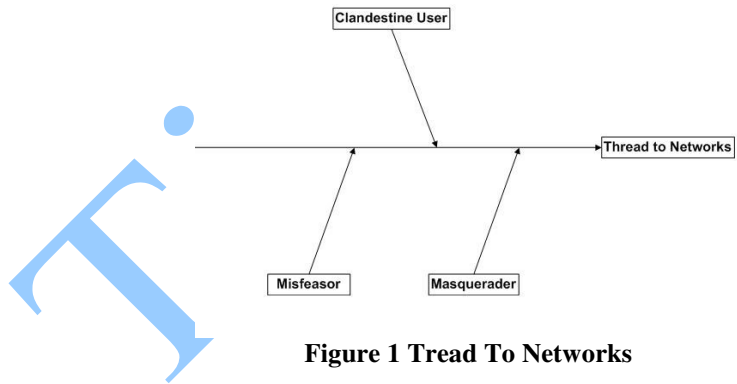

**Figure 1 Tread To Networks**

In an unprotected network environment any client can apply to any server for service. This obvious security risk is that of impersonation. An opponent can pretend to be another client and obtain unauthorized privileges on server machine. To counter this threat, servers must be able to confirm the identities of clients who request services.

---

[1] Kerberos is an authentication service developed as part of Project Athena at MIT
[2] Kerberos Version 1 thought 3 were development versions. Version 4 is the original Kerberos

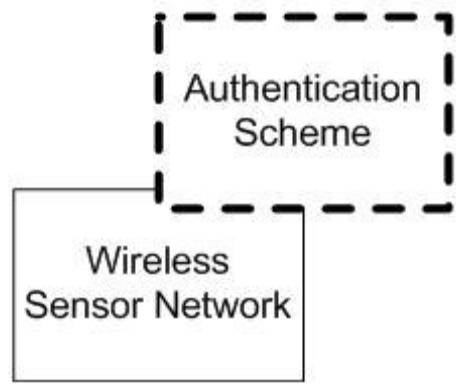

**Figure 2 Authentication Layer in WSN**

Verification of the user in the wireless sensor network we have added the authentication scheme layer above the wireless sensor network which authorized the user to access the wireless sensor network without verification the user can't access the wireless sensor network. The authentication scheme is based on the Kerberos server authentication scheme. The detail of the scheme is explained in the section 3.

The rest of the paper is structured as following first we give the overview of the authentication techniques next we discuss the overview of the traffic in the Wireless sensor network. Then we discuss the Kerberos Server Architecture and its components and then their components requirement for setting the Kerberos environments finally we conclude the paper.

# 1. Background

We have use the term wireless sensor network to refer to a heterogeneous system combining tiny sensors and actuators with general-purpose computing elements. Sensor networks may consist of hundreds or thousands of low-power, low-cost nodes, possibly mobile but more likely at fixed locations, deployed en masse to monitor and affect the environment. For the remainder of this paper we assume that all nodes locations are fixed for the duration of their lifetime.

Authentication server techniques - that knows the passwords of all the users and stores these in a centralized database. In addition, the authentication server shares a unique secret key with each server. These keys have been distributed physical or in come other secure manner.

X.509 authentication service is a part of X.500 series of recommendation that define a directory service. The directory id, in effect, a server or distributed set of server that maintains a database of information about users.X.509 defines a framework for the provision of authentication services by the X.500 directory to its users.X.509 also define an authentication protocol based on the use of public key certificates.

Digital signatures can be used in for the authentication assuming that each sensor node in the network in pre loaded with the public key of some certification authority [KSW04].

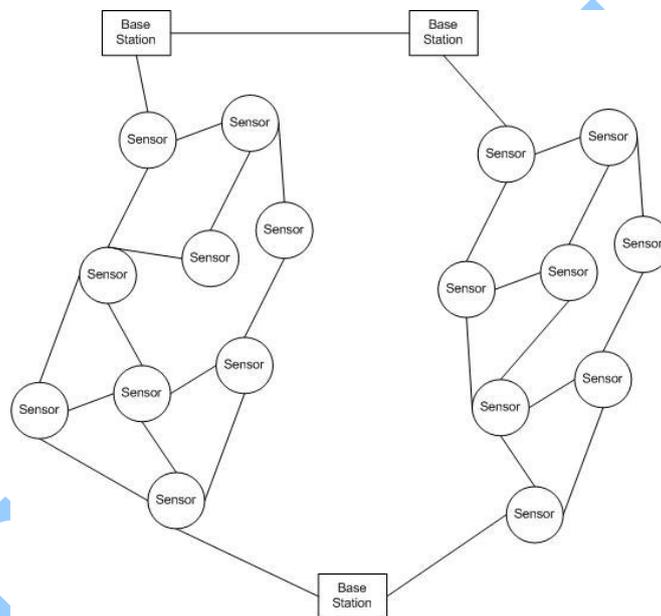

**Figure 3 A representative Wireless Sensor Network Architecture**

Base stations are typically many orders of magnitude more powerful than sensor nodes. They might have workstation or laptop-class processors, memory, and storage, AC power, and high-band- width links for communication amongst themselves. However, sensors are constrained to use lower-power, lower-bandwidth, shorter range radios, and so it is envisioned that the sensor nodes would form a multi hop wireless network to allow sensors to communicate to the nearest base station.

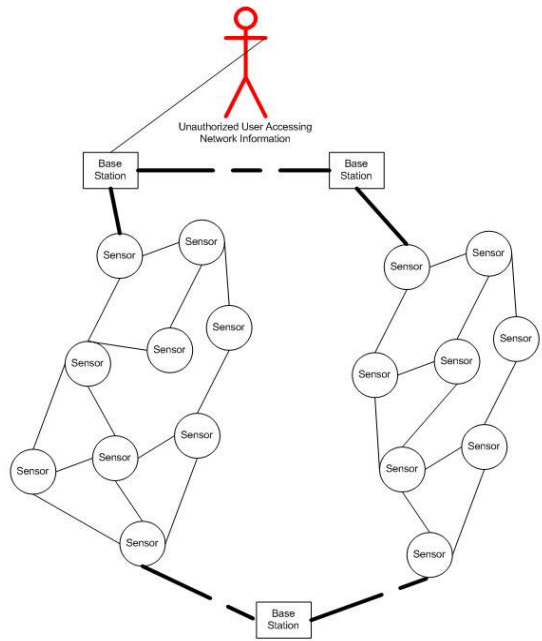

**Figure 4 Unauthorized User Accessing the Base Station information**

The figure 4 explain that as there is no authentication scheme implemented in the base station so the unauthorized user is access the bases station because there is no verification of the user.

## 2. Traffic in Wireless Sensor Network

Traffic in sensor networks can be classified into one of three categories:

*A. Many-to-one*

Multiple sensor nodes send sensor readings to a base station or aggregation point in the network.

*B. One-to-many*

A single node (typically a base station) multicasts or floods a query or control information to several sensor nodes.

*C. Local communication*

Neighboring nodes send localized messages to discover and coordinate with each other. A node may broadcast messages intended to be received by

71

all neighboring nodes or unicast messages intended for a only single neighbor intended for only single neighbor[3].

## 3. Kerberos Server Architecture

There are two main components of Kerberos servers
- Authentication Server
- Ticket Granting Server

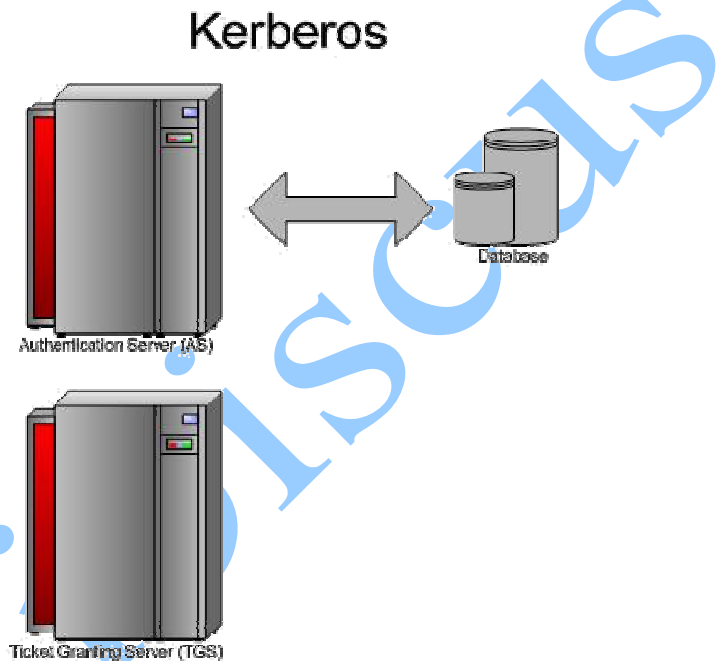

**Figure 5 Kerberos Architecture**

### A. *Authentication Server*

Authentication server knows the password of all the users and stores these in a centralized database. The authentication server shares a unique secret key with each server. These keys have been distributed to the user in some secure manners.

### B. *Ticket Granting Server*

Ticket granting server issues tickets to users who have been authenticated to authentication server. Then the user first requests a ticket from the authentication server. This ticket is saved by the user. Each time the user authenticate itself the ticket granting server then grants a ticket for the particular server/Base Station. The user save each service granting ticket and uses it to authenticate its user to a server each time a particular service is requested.

---

[3] By neighbor we mean a node within normal radio range.

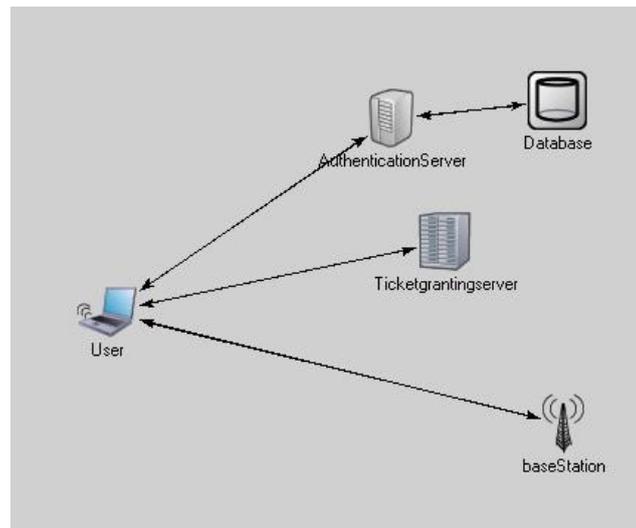

**Figure 6 Overview of Kerberos**

- Client requests a ticket granting ticket on behalf of the user by sending its users ID to the authentication Server.
- The authentication server responds with a ticket that is encrypted with a key. When the ticket arrives at the client, the client prompts the user for the password, generate the required and decrypt the incoming message.
- The client requests the service-granting ticket on behalf of the user. Then client transmit a message to the Ticket granting ticket containing the users ID and the ID of the desired service, and the ticket granting ticket.
- The ticket granting server verifies the ticket it checks that the time limit has not expired. Then the ticket granting ticket issues a ticket to grant access to the requested service.
- The client requests access to a service on behalf of the user. For this purpose the client transmits a message to the server containing the user ID and the service granting ticket. The server authenticates by using the contents of the tickets

73

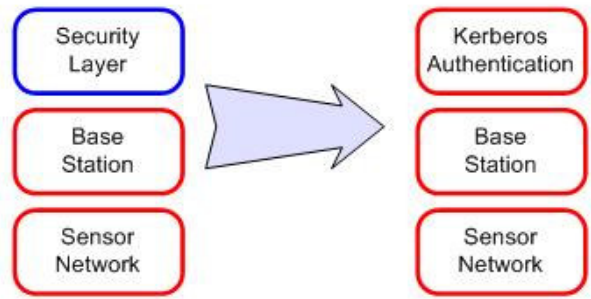

**Figure 7 Adding New layer in Wireless Sensor Network**

In fig 7 we added the new layer in the wireless sensor architecture. The sensor network transfer the information to the base station the based station is directly connected to the user or the middleware and in the pervious architecture no security layer is present so there is always a demand of security layer in the wireless sensor network. So in this paper we presented a Kerberos authentication scheme to protect the wireless sensor network for the unauthorized user. By adding the security layer in the wireless sensor network we can prevent the network from the different security thread. The base station can't be accessed directly because the user has to authentication him/her from the Kerberos server and then obtain the ticker to access the base station to retrieve the in the information provided by the sensors.

## 4. Kerberos Environment (Compoments and Requirements)

The Kerberos environment consist of the following components:
- Kerberos Server
- Number of Clients
- Number of Application Server/Base Station

The Kerberos server must share a secret key with each server.

The Kerberos server must have user ID and hasted password of all participating users in its database.

The Kerberos server in each interoperating realm shares a secret key with the server in the other realm.

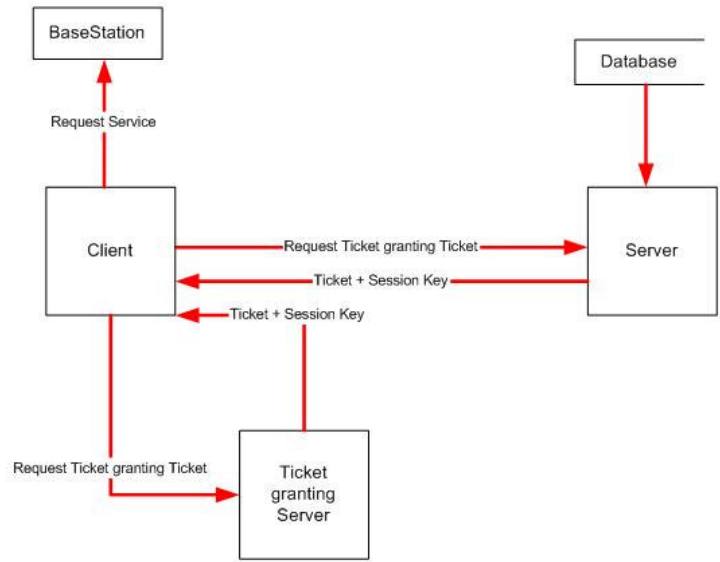

**Figure 8 Dataflow**

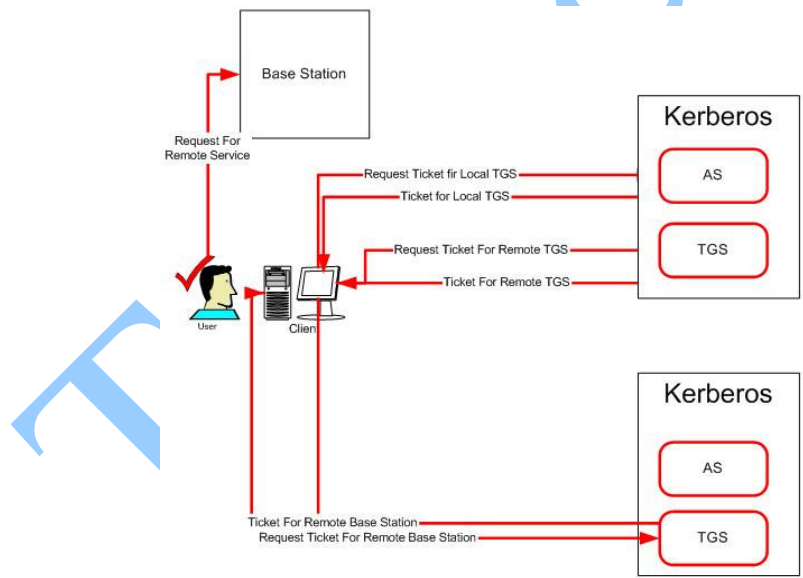

**Figure 9 Request for base station request in another realm**

The client requests the ticket from the authentication server. If the client is authenticated the server provide the ticket to the client. The client decrypts the ticket using it password. Then client transmit a message to the Ticket granting ticket containing the users ID and the ID of the desired service, and the ticket granting ticket. The ticket granting server verifies the

75

ticket it checks that the time limit has not expired. Then the ticket granting ticket issues a ticket to grant access to the requested service. The client transmits a message to the Base station of the wireless sensor network containing the user ID and the service granting ticket. The base station server authenticates by using the contents of the ticket.

## 5. Authentication as Energy Efficient System for WSN

The proposed system in this paper not only helps to secure the wireless sensor network but it is help to improve the lifetime of the base station of the wireless sensor network.

With the help of Kerberos authentication system only those user retrieve the information from the base station that are authorized. So the base station doesn't send unnecessary packet. This help the bas station to save energy and helps to increased to overall base station time

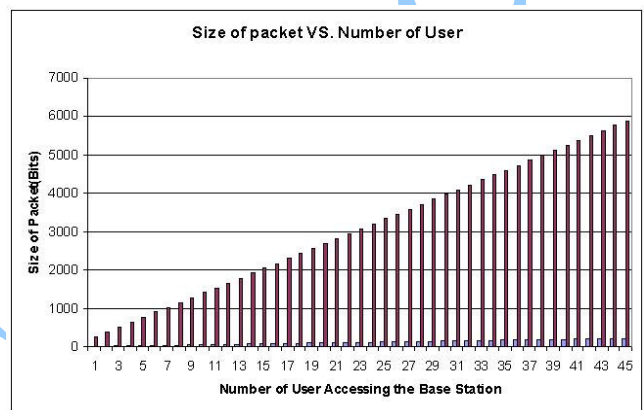

**Figure 10 Relationship between Sizes of Packet VS. Number of Users**

In fig 10 the number of the user accessing the base station of the wireless sensor network are consider on the x-axis and the size of the packet are consider on the y-axis as we can observe from the figure that as the user accessing the base station increased the size of the packet on the network as increased. So from the results we obtain that the user accessing the network is directly proportional to the size of the packets.

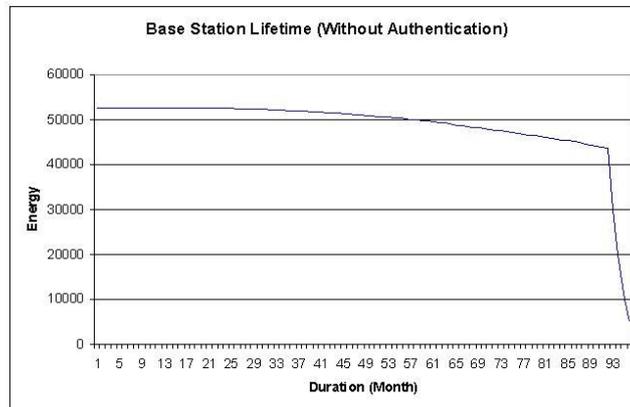

**Figure 11 Base station Lifetime without Authentication**

In fig 11 the duration of the lifetime of the base station is considered on the x-axis of the graph and the energy is considered on the y-axis of the graph. This result is generated for the system where no authentication was performed we observe from the graph that the lifetime is less and the energy of the base station start to decreased quickly because it is sending message to all the user without verification.

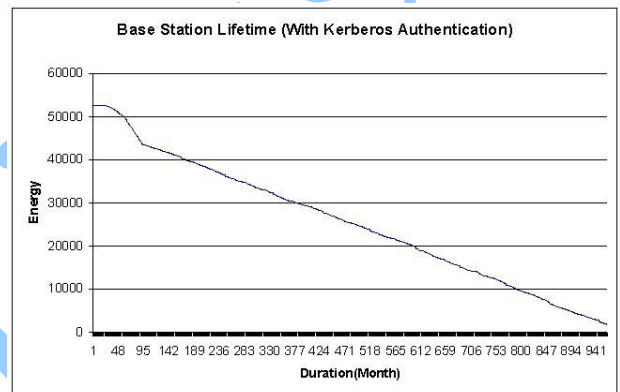

**Figure 12 Base station lifetime with Kerberos authentication**

In fig 12 the duration of the lifetime of the base station is considered on the x-axis of the graph and the energy is considered on the y-axis of the graph. This result is generated for the system where authentication is performed with the help of Kerberos server authentication. As in authentication the base station sends the message to those users who are authenticated and this help the base station to save the energy and by saving the energy the lifetime of the network increased.

77

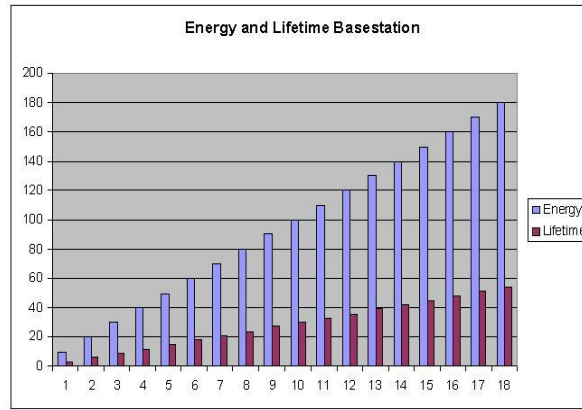

**Figure 13 Relationship between Energy and Lifetime**

In fig 13 we observe that the base station energy is directly proportional to the life time. As it can be observed from the above figure that as the energy increased the lifetime of the base station also increased and the energy decreased the lifetime of the base station also decreased. As it can be calculate that they are directly proportional to each other.

## Conclusions

The main purpose of the paper is to provide an overview of the architecture of the Kerberos authentication service and using the Kerberos authentication services in Wireless sensor network to secure the base station from unauthorized user. The proposed system in this paper not only helps to secure the wireless sensor network but it is help to improve the lifetime of the base station of the wireless sensor network.

We conclude that the wireless Sensor networks are increasing day by day we must implemented all types of security policies in our wireless sensor network security system. To protect our sensor network from the security risk a better solution is to user Kerberos authentication services.

## References

[G+04]   N. Gura, A. Patel, A. Wander, H. Eberle and S. C. Shantz - *Comparing elliptic curve cryptography and RSA on 8-bit CPUs*,